\documentclass[conference]{IEEEtran}
\IEEEoverridecommandlockouts
\usepackage{graphicx}
\usepackage{xcolor}
\usepackage{tikz}
\usepackage{gensymb}
\usetikzlibrary{positioning}

\usepackage{bm}
\usepackage{amssymb}
\usepackage{amsmath}
\usepackage{siunitx}
\usepackage{subfigure}

\usepackage{hyperref}
\hypersetup{
    colorlinks=true,
    linkcolor=blue,
    filecolor=magenta,      
    urlcolor=cyan,
    pdftitle={Overleaf Example},
    pdfpagemode=FullScreen,
    }

\begin{document}
%
\title{Dynamic Real-Time Ambisonics Order Adaptation for Immersive Networked Music Performances}
%
%
%
\author{
\IEEEauthorblockN{Paolo Ostan$^1$, Carlo Centofanti$^2$, Mirco Pezzoli$^1$, Alberto Bernardini$^1$, Claudia Rinaldi$^3$, Fabio Antonacci$^1$}\\
\IEEEauthorblockA{$^1$Dipartimento di Elettronica, Informazione e Bioingengeria, Politecnico di Milano, Italy \\ 
$^2$ Department of Information Engineering, Computer Science, and Mathematics, University of L'Aquila, Italy\\
$^3$ CNIT- National Inter-University Consortium for Telecommunications, Research Unit of the University of L'Aquila, Italy }
\thanks{This work was partially supported by the European Union - Next Generation EU under the Italian National Recovery and Resilience Plan (NRRP), Mission 4, Component 2, Investment 1.3, CUP D43C22003080001, partnership on ``Telecommunications of the Future'' (PE00000001 - program ``RESTART''), and by ``REPERTORIUM project. Grant agreement number 101095065. Horizon Europe. Cluster II. Culture, Creativity and Inclusive Society. Call HORIZON-CL2-2022-HERITAGE-01-02.''.
}}

\maketitle
\begin{abstract}
Advanced remote applications such as Networked Music Performance (NMP) require solutions to guarantee immersive real-world-like interaction among users.
Therefore, the adoption of spatial audio formats, such as Ambisonics, is fundamental to let the user experience an immersive acoustic scene. 
The accuracy of the sound scene reproduction increases with the order of the Ambisonics enconding, resulting in an improved immersivity at the cost of a greater number of audio channels, which in turn escalates both bandwidth requirements and susceptibility to network impairments (e.g., latency, jitter, and packet loss). These factors pose a significant challenge for interactive music sessions, which demand high spatial fidelity and low end-to-end delay.
We propose a real-time adaptive higher-order Ambisonics strategy that continuously monitors network throughput and dynamically scales the Ambisonics order. 
When available bandwidth drops below a preset threshold, the order is lowered to prevent audio dropouts; it then reverts to higher orders once conditions recover, thus balancing immersion and reliability. A MUSHRA-based evaluation indicates that this adaptive approach is promising to guarantee user experience in bandwidth-limited NMP scenarios.
\end{abstract}
\begin{IEEEkeywords}
higher order ambisonics, spherical harmonics, network protocol, networked music performance
\end{IEEEkeywords}
%
\IEEEpeerreviewmaketitle
\section{Introduction}
\IEEEPARstart{N}{etworked} Music Performance (NMP) has emerged as a significant area of scholarly inquiry within telepresence and collaborative artistic practices. It facilitates real-time musical interactions and shared audience experiences across geographical distances \cite{rottondi2016NMP}. The COVID-19 pandemic has further highlighted the importance of NMP, forcing musicians to explore online collaboration methods \cite{Loveridge2020NetworkedMP}. As the field advances, Immersive NMP (INMP) is gaining increasing attention. 
INMP aims to enhance the sense of presence and engagement by preserving not only temporal and spectral fidelity but also the crucial spatial and acoustic cues of musical signals \cite{turchet2024INMP}. This heightened immersion is posited to foster stronger interpersonal connections among performers and listeners, effectively mitigating geographical limitations and enriching artistic collaborations.
Ambisonics technology has surfaced as a prominent methodology for achieving immersive spatial audio within networked environments \cite{miotello2024homula}. 
Ambisonics offers a playback-agnostic encoding of the sound field, enabling versatile and high-fidelity reproduction of three-dimensional auditory scenes \cite{bookAmbi2019}. In contrast to conventional multi-channel systems predicated on fixed loudspeaker configurations, Ambisonics provides inherent scalability and compatibility across diverse rendering modalities—ranging from binaural reproduction via headphones to complex acoustic dome installations. Furthermore, Higher-Order Ambisonics (HOA) facilitates the capture of finer spatial details, thereby improving the precision in localizing and conveying the directionality of sound sources. In \cite{Gurevich2011AmbisonicSF} authors have demonstrated the viability and effectiveness of HOA for real-time spatialization in networked music performance, highlighting its creative potential in exploring space and distance in collaborative music.
However, real-time transmission of HOA signals over heterogeneous networks characterized by variable bandwidth and latency presents considerable challenges. Fluctuations in network throughput, latency, and packet loss can negatively affect temporal coherence and perceptual quality essential for high-fidelity NMP \cite{rottondi2016NMP,Mezza:2023HybridPLC}. 

To address these network-induced impairments, an adaptive strategy based on the dynamic adjustment of the Ambisonics order in response to real-time fluctuations in network conditions is proposed. This adaptive approach is designed to optimize the immersive auditory experience while minimizing disruptive artifacts such as audio dropouts or excessive latency. Mróz et al. \cite{mroz2023commonly} recently presented a toolchain for live streaming music events with HOA audio and 4k 360 vision, emphasizing the growing interest in immersive live streaming.
Adaptive streaming methodologies are well-established in the video domain (e.g., MPEG-DASH \cite{Thang2012DASH}), where dynamic bitrate adjustments are employed to prevent buffering. 
Similarly, in the audio domain, real-time data reduction and compression techniques, including adaptive codecs, are prevalent \cite{namazi2022spatial}. 
However, research specifically addressing spatial audio adaptation remains limited \cite{namazi2022spatial}, and to the best of our knowledge, no prior work has thoroughly investigated real-time Ambisonics order adaptation within NMP contexts, particularly in scenarios exhibiting significant bandwidth and latency variability.
In this study, we investigate the feasibility of real-time Ambisonics order adaptation for INMP. 
 Our investigation encompasses simulating encoding and decoding strategies that can smoothly transition among different Ambisonics orders without introducing audible discontinuities, and conducting subjective listening tests (via MUSHRA) to evaluate the perceived impact of instant vs. cross-faded order transitions on audio quality and immersion. 
\section{Scenario}%
\begin{figure}%
    \centering%
    \includegraphics[width=0.9\linewidth]{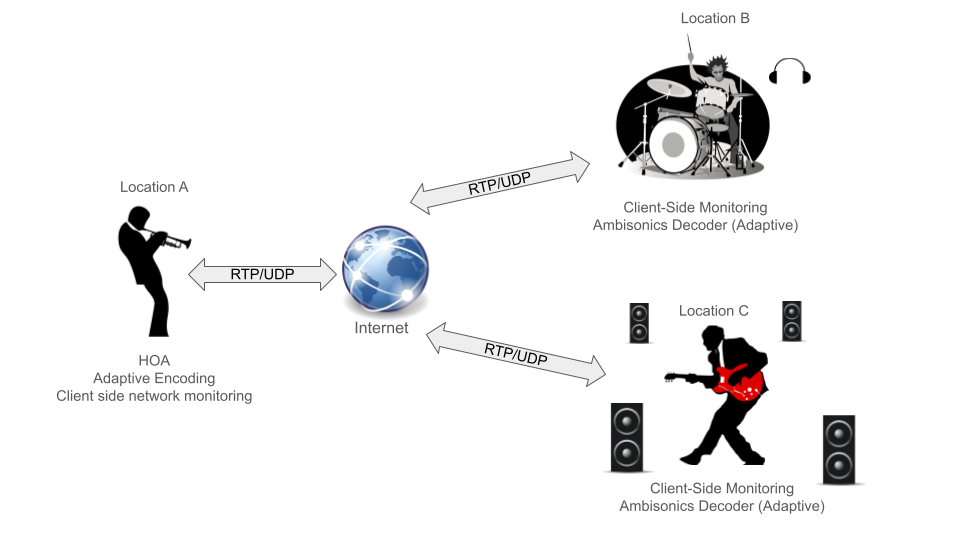}
    \caption{Schematic of adaptive-order Ambisonics in NMP. Location A: primary source capture and adaptive encoding. Locations B/C: remote clients with adaptive decoding. Real-time network monitoring dynamically adjusts Ambisonics order over RTP/UDP.}\label{fig:scenario}\vspace{-1.5em}%
\end{figure}
To contextualize our investigation, we posit a scenario of a networked musical jam session involving musicians situated in geographically disparate locations (see Fig.~\ref{fig:scenario}).  
We consider three musicians, located at Location A, Location B, and Location C.  The objective is to enable real-time co-performance, facilitating an immersive auditory experience that transcends conventional stereophonic reproduction and is suitable for various musical ensembles and spatial audio setups \cite{tomasetti2023immersive}.
\vspace{-0.2em}
\subsection{Capture Stage}
At Location A, a primary acoustic instrument or sound source is captured utilizing a higher order Ambisonics microphone array (HOM) of order $N$, strategically positioned at $\bm{r} \in \mathbb{R}^3$ to capture the acoustic field. 
This includes both the spatial characteristics of the instrument and the inherent room acoustics.
The ambisonics coefficients $\alpha_{n,m}(t)$ are obtained from the raw signals of the $Q$ capsules as
\begin{equation}\label{eq:ambisonics signals}
    \alpha_{n,m}(t) = g_{n}\sum_{q=1}^Q w_q p(\bm{r}'_q, t)Y_{n,m}(\theta_q,\phi_q),
\end{equation}
where $t \in \mathbb{Z}$ is the discrete-time index, $n=0, 1,\dots, N-1$ and $m = -n,\dots,n,\, \forall n$ are the order and degree of the encoding, respectively, $p(\bm{r}'_q, t)$ is the sound pressure at the $q$th sensor, $\theta_q$  and $\phi_q$ are the azimuth and elevation positions of the $q$th sensor, $g_n$ is the $n$th order normalization factor and $w_q$ are weights that ensure the spherical harmonics decomposition validity \cite{rafaely2015fundamentals}. 
The number of channels $Q$ in the array determines the maximum order $N$ such that $Q\geq(N+1)^2$, while the spherical harmonics bases in \eqref{eq:ambisonics signals} are defined as
\begin{equation} \label{SH}
    Y_{n,m}(\theta_q, \phi_q) = \sqrt{\frac{(2n+1)}{4\pi} \frac{(n-m)!}{(n+m)!}} P_{n}^{m}(\cos\theta_q) e^{im\phi_q},
\end{equation}
with $P_{n}^{m}(\cdot)$ the real-valued associated Legendre function.

The acquired acoustic field at each discrete-time instant can be represented as a collection of encoded SH-based coefficients:
\begin{equation}\label{eq:vector_ambisonics}
        \mathbf{A}(t) = [\alpha_{0,0}(t), \alpha_{1,-1}(t), \alpha_{1,0}(t),  \alpha_{1,1}(t),  \dots ,\alpha_{N,N}(t)]^T, 
\end{equation}
where $\mathbf{A}(t) \in \mathbb{R}^{C}$ and $C=(N+1)^2$ is the number of Ambisonics channels depending on the order $N$. 
It is worth noting that the maximum number of SH-based coefficients depends on the order of the HOM according to \eqref{eq:ambisonics signals}, thus $\mathbf{A}(t)$ has length $C \leq Q$.

At Location B and Location C, secondary instruments or sound sources are acquired via conventional microphones or direct line inputs. For simplicity in this scenario, we assume that these signals also undergo pre-processing to facilitate their integration into a spatialized auditory environment, potentially using spatial panning \cite{coleman2017object} or upsampling techniques \cite{miotello2024physics}. 
This setup allows for a central, spatially rich sound source to be adaptively transmitted alongside other contributing instruments in the networked performance.
\vspace{-0.5em}
\subsection{Communication protocol}
The resultant audio streams originating from Location A, Location B, and Location C are transmitted through a public IP network infrastructure. 
The signal is segmented into discrete time frames with frame-length $L_F$, expressed in matrix form as: $\mathbf{F}_k = [\mathbf{A}(t_{k}),\   \mathbf{A}(t_{k} + 1),  \dots ,     \mathbf{A}(t_{k} + L_F-1) ]$
 where $t_k$ defines the index referring to the first sample of the time window covered in packet $k$. 
 To ensure low-latency real-time transmission, the Real-time Transport Protocol (RTP) over the User Datagram Protocol (UDP) is used, as low latency is critical for effective musical interaction \cite{Loveridge2020NetworkedMP}.  
Each frame $\mathbf{F}_k$ is encapsulated into a UDP payload of size $L_P$, where
\begin{equation}\label{eq:packet_size}
    L_P =  \sum_{n=0}^{N-1} \sum_{m=-n}^{n} B(n,m) \cdot L_F,
\end{equation}
with $B(n,m)$ denoting the bit-depth allocated for each coefficient $\alpha_{n,m}(t)$.
For transmission, each UDP packet at time $t_k$ is streamed over a network using an RTP-based encapsulation:

\begin{equation}\label{eq:stream}
\mathbf{S}_k = \mathcal{E} (\mathbf{F}_k), \end{equation}
where the operator $\mathcal{E}$ encapsulates the frame $\mathbf{F_k}$ assigning an header containing metadata such as sequence numbers, timestamps, and optional error correction bits to mitigate packet loss. 
On the receiving end, the decoding process is structured as
\begin{equation} \label{eq:packet_decoding}
    \hat{\mathbf{F}}_k = \mathcal{D}(\hat{\mathbf{S}}_k),    
\end{equation}
where $\mathcal{D}$ represents the decoding function that extracts from the received stream $\hat{\mathbf{S}}_k$ the frame $\hat{\mathbf{F}}_k$ containing the HOA coefficients and applies jitter buffering.
These signals are then decoded for binaural or loudspeaker rendering using an inverse spherical harmonic transform \cite{rafaely2015fundamentals}.

Real-time monitoring of network conditions is paramount and may be implemented via a distributed architecture, encompassing both server-side and client-side components. Consequently, each musician’s workstation not only serves its own audio stream but also actively monitors network metrics in real time. On the server side, the software monitors downstream bandwidth, latency (via RTP timestamps), and packet loss (from RTP drops). These measurements are fed into the adaptive control module, allowing for timely adjustments to preserve audio quality under fluctuating conditions. Simultaneously, each workstation functions as a client, decoding streams from other locations, and measuring the same suite of metrics from the receiver side. By combining server-side and client-side observations, a more comprehensive view of the network state is achieved, facilitating robust, low-latency performance across the public IP infrastructure.
Whenever the server-side module decides to adjust the Ambisonics order due to network impairments, metadata embedded within RTP packets inform client-side decoders of the new encoding configuration, thus enabling seamless adaptation without introducing audible discontinuities.
\vspace{-0.5em}
\subsection{Communication network}
Modern IP networks can experience various performance impairments such as latency, jitter, and packet loss \cite{tanenbaum2013computer}. In this work, we focus specifically on bandwidth saturation, that is, the condition in which a link capacity is fully utilized. 

Under saturation conditions, network buffers experience queue overflow, leading to increased packet loss and transmission delays.
In case of saturation, queues on routers can overflow, causing packet drops and higher latency. Since UDP lacks built-in congestion control, these conditions often exacerbate network inefficiencies. Consequently, real-time adaptation of the Ambisonics order (Fig.~\ref{fig:scenario}) becomes important to maintaining acceptable audio quality under fluctuating network loads.
To mitigate such inefficiencies, we propose to dynamically adapt the Ambisonics order based on real-time network conditions as detailed in the next section. 
\vspace{-0.5em}
\section{Ambisonics Order Adaptation Mechanism}
The core of the proposed system is the adaptive Ambisonics order control module. 
\subsection{Adaptation Triggering Logic}
The adaptation system dynamically adjusts the Ambisonics order $N$ based on real-time network conditions to ensure seamless transmission while minimizing latency. The server continuously monitors the available bandwidth
\begin{equation} 
C_A = \frac{L_R}{T_R}, 
\end{equation}
where $L_R$ is the total received packet load within a measurement interval $T_R$.
Currently, a reduction in Ambisonics order is triggered solely when the server-side estimated available bandwidth falls below a predefined configurable threshold, e.g., $\SI{2}{\mega bps}$. 
This threshold can be adjusted according to the complexity of audio signals and the latency constraints of specific applications.
\vspace{-0.5em}
\subsection{Adaptive Control Mechanism}
Upon trigger activation, the Ambisonics encoder at Location A (primary sound source) dynamically reduces the encoding order. This is achieved by transitioning from $N$th order (e.g., third order) to lower orders (e.g., second or first order Ambisonics), and potentially further down to order zero if network conditions continue to degrade.  
This approach aligns with adaptive streaming strategies used for the delivery of HOA content \cite{rudzki2021hoast}. 
Concurrently, clients are programmatically notified of the order transition via RTP metadata embedded within the audio stream.
To maintain an acceptable transmission rate, the Ambisonics order is dynamically reduced to ensure that the packet load remains within network capacity. 
Specifically, given a target bitrate constraint $R_{\max}$, the order adaptation strategy for the $k$th packet follows 
\begin{equation} 
    N'_{k} = \max \left\{ N \mid L_P(N) \leq R_{\max} \cdot T_P \right\},
\end{equation}
$N'_{k} \in [0, N]$ is the adjusted Ambisonics order leading to a channel count $C=(N'_{k}+1)^2$ in \eqref{eq:vector_ambisonics}, $L_P(N)$ \eqref{eq:packet_size} is the packet load size for a given order $N$ and $T_P$ is the packet transmission interval.
This strategy ensures that the transmission remains within bandwidth constraints while preserving the highest possible spatial resolution $N'_{k}$.
\vspace{-0.5em}
\subsection{Receiver-Side Adaptive Decoder}
At the receiving locations (Location B and Location C, and potentially remote audience listening posts), Ambisonics decoders are notified of the update of the decoding order $N'_{k}$ to facilitate seamless order transitions across varying Ambisonics orders. 
Additionaly, after the server-side requests the reduction of the Ambisonics order, a fade-out strategy can be applied to the high-order channels that will not be transmitted after the reduction is applied. 
Instead of an instantaneous drop in order, the energy of the higher-order Ambisonics coefficients is gradually attenuated before they are removed from transmission.
This approach is considered to minimize discontinuities and prevent spatial degradation that could affect user experience.
After the decoding in \eqref{eq:packet_decoding} is computed, the audio buffer is updated as 
\begin{equation} 
    \hat{\mathbf{A}} (t) = \hat{\mathbf{A}} (t,N'_{t_k}) +  w_{t} \mathbf{H}(t, N_{t_k}),
\end{equation}
where $\hat{\mathbf{A}} (t,N'_{k})$ represents the decoded Ambisonics signal extracted from $\hat{\mathbf{F}}_k$, possibly at the reduced order $N'_{k}$, $\mathbf{H}(t, N_{k})$ denotes the set of high-order Ambisonics coefficients that are being faded out, defined as 
\begin{equation}
    \mathbf{H}(t, N_{k}) = \hat{\mathbf{A}} (t,N_{k}) - \hat{\mathbf{A}} (t,N'_{k}),
\end{equation}
with $w_{t}$ a time-dependent fade-out function that gradually attenuates the contribution of the high-order components over a predefined duration $T_f$.
The choice of $T_f$ significantly impacts both perceptual quality and resilience to network fluctuations. 
A longer $T_f$ provides a smoother spatial transition but extends the period where the system is vulnerable to packet loss due to bandwidth saturation, as the network may struggle to transmit the amount of data necessary to complete the fade-out process.
On the other side, a short or null cross-fade allows for instantaneous adaptation of the spatial resolution.
The output of the adaptive decoder is then routed to binaural headphones or multi-channel loudspeaker arrays for immersive sound reproduction.
\vspace{-0.5em}
\begin{figure*}[tbh]
    \centering
    \begin{subfigure}[]{\includegraphics[width=0.72\columnwidth]{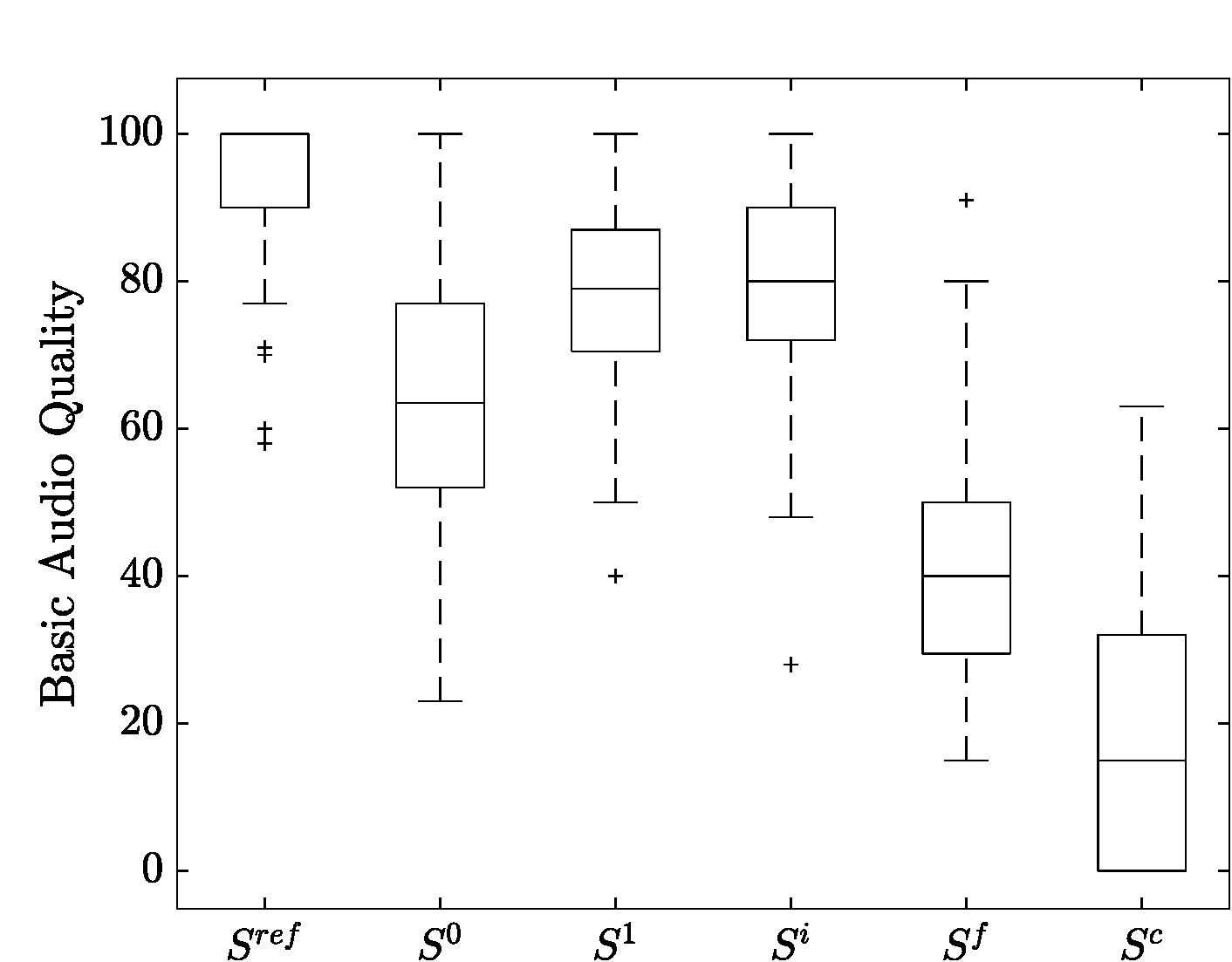}\label{fig:baq}}%
    \end{subfigure}%
    \begin{subfigure}[]{
    \includegraphics[width=0.72\columnwidth]{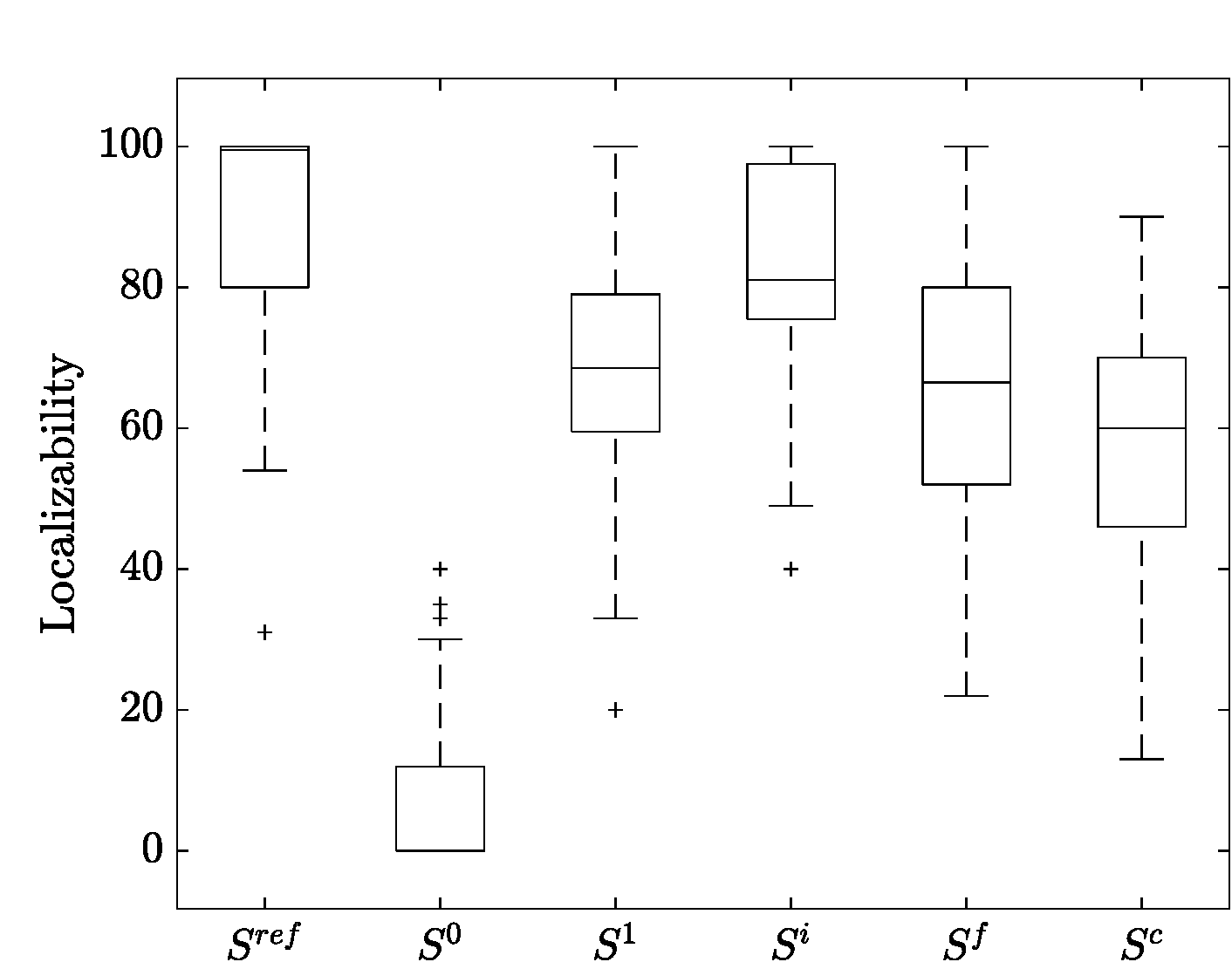}\label{fig:localizability}}%
    \end{subfigure}%
    \caption{(a) Basic Audio Quality and (b) Localizability scores aggregated for every stimulus evaluated in the perceptual test.}\vspace{-1em}
\end{figure*}
\section{Evaluation}
To assess the perceptual impact of the dynamic order adaptation system, a Multiple Stimuli with Hidden Reference and Anchor (MUSHRA) test was conducted \cite{schoeffler2015towards}. Participants listened to a series of spatial audio stimuli presented over headphones using binaural rendering and were asked to evaluate the Basic Audio Quality (BAQ), \cite{schoeffler2016evaluation}, and the localizability, \cite{immersion2019}, of each stimulus on a scale ranging from \textit{0 - bad} to \textit{100 - excellent}.
\subsection{Perceptual test setup}
 The test included a third-order Ambisonics ($N=3$) reference and four degraded versions of the stream $\mathbf{S}_k$ \eqref{eq:stream} to evaluate the perceptual differences introduced by network impairments and adaptation strategies. 
 In particular, the test comprised six distinct stimuli, each serving a specific evaluative purpose:
 \begin{itemize}
     \item explicit reference: third-order audio stream;
     \item hidden reference $\mathbf{S}^{\text{ref}}$: third-order audio stream presented without explicit identification to determine evaluators reliability;
     \item low spatial quality solution $\mathbf{S}^{0}$: omnidirectional audio stream, solution for minimum bandwidth consumption while preserving basic audio quality and sacrificing immersivity, eliminating directional cues;
     \item safe policy solution $\mathbf{S}^{1}$: a continuously maintained first-order stream that is presented throughout the entire duration of the audio sample, ensuring a stable and balanced quality level and spatiality;
     \item instantaneous adaptation $\mathbf{S}^{\text{i}}$: the stimulus starts as a third-order stream that is instantaneously reduced to first order; this strategy is optimized to enhance audio quality by transmitting all packets under normal conditions while dynamically lowering the order during network fluctuations to ensure consistent delivery;
     \item cross-faded adaptation $\mathbf{S}^{\text{f}}$: this stimulus involves reducing the third-order stream to first order, but also incorporating a one second cross-fade; this approach minimizes audio quality differences during transitions but remains vulnerable to packet loss during the transition period;
     \item corrupted transmission $\mathbf{S}^{\text{c}}$: this stimulus is generated from a third-order stream that is transmitted with induced packet loss; it provides a middle-ground quality level against which the proposed solutions are compared.
 \end{itemize}
A total of $6$ sound scenes has been evaluated considering different source signals from \cite{cd2008sound}. 
Each sound scene considers two acoustic sources; one static source and one dynamically moving within the auditory space. This setup allows for the evalutation of the perceptual characteristics of the reproduced sound scene under different conditions.
Two different scenarios were analyzed. 
In the first scenario, the moving source maintains a fixed elevation angle while its azimuth ranges from $\SI{-90}{\degree}$ to $\SI{90}{\degree} $, is used to evaluate the horizontal localizability. In the second scenario, the source remains at a fixed azimuth, while its elevation gradually changes from $\SI{0}{\degree}$ to $\SI{180}{\degree}$, allowing for an assessment of the system’s ability to preserve vertical localization cues.
Furthermore, the generated $6$ sound scenes were processed to obtain each of the six stimuli described previously. 
In particular, a $\SI{5}{\percent}$ packet loss is simulated due to bandwith constraints. 
Order reduction was applied dynamically on the basis of the proposed adaptation strategies.
All sound scenes were rendered in binaural format using Head-Related Transfer Functions (HRTFs) \cite{lindau2007binaural} converted into the spherical harmonics (SH) domain using the Magnitude Least Squares (MagLS) method \cite{schorkhuber2018binaural}. This approach ensures a perceptually optimized reconstruction of spatial cues while maintaining consistency across Ambisonics decoding in the case of listening tests through headphones. 

The perceptual evaluation of the transmitted spatial audio is based on the assessment of two perceptual attributes.
BAQ evaluates the overall fidelity of the reproduced sound, taking into account aspects such as timbral accuracy, presence of distortion or artifacts.
Localizability measures the assessors' ability to evaluate the spatial extent and location of sound sources in the reproduced soundfield  \cite{lindau2014spatial}.
\subsection{Results}
11 participants (10 men and 1 woman) between 27 and 46 years of age, mainly with audio engineering background, were asked to perform the online listening test. 
On average, the test required 15 minutes to be completed. 

The results on BAQ are reported in Fig.~\ref{fig:baq}.
We notice that, as expected, the corrupted stream $\mathbf{S}^{\text{c}}$ received the lowest evaluations among the participants due to the presence of discontinuities in the stream.
The adoption of the proposed adaptation strategy provides improved overall audio quality. 
Although the scores registered for long-fading adaptation $\mathbf{S}^{\text{f}}$ are far from the hidden reference, it still shows an improvement with respect to the corrupted signal. 
This suggests that a long cross-fading solution might retain audible artifacts for the rate of packet loss considered.
Differently, the instantaneous adaptation $\mathbf{S}_{\text{i}}$ achieves a performance that is statistically comparable with the safe policy $\mathbf{S}^1$ and close to the reference, suggesting that an instantaneous adjustment of the Ambisonics order does not significantly impact the audio quality. 
The omnidirectional stream $\mathbf{S}^{0}$, instead, obtains scores that are in the middle between the reference and the corrupted stimulus. 
In fact, although no packet loss is present for $\mathbf{S}^{0}$, differences in the tone are perceivable with respect to the reference due to the missing higher-order channels in the stream.

In Fig.~\ref{fig:localizability} the localizability results are reported.
As one might expect, the omnidirectional stimulus $\mathbf{S}^{0}$ obtains the lowest scores due to the lack of spatial characteristics. 
At the same time, the other adaptation strategies achieved higher scores. 
In particular, $\mathbf{S}^{\text{i}}$ achieves the best result, being in line with the reference scores. 
As far as  $\mathbf{S}^{\text{f}}$ is concerned, the evaluations are in line with a first-order HOA safe policy.
On the one hand, the results suggest that a slower adaptation of the order might reduce the localizability approaching the perceptual capabilities of a low-order Ambisonics. 
On the other hand, the corrupted stream, although provided at the highest spatial resolution, reduces the ability of the assessors to localize the acoustic sources under test.
In combination with BAQ scores $\mathbf{S}^{\text{i}}$ confirms to have the best adaptation strategy in order to guarantee both tone quality and high spatial characteristics.
This confirms the potentiality of the proposed adaptation strategy in improving the overall immersive experience in the context of INMP. 
\vspace{-0.5em}
\section{Conclusions and Future Work}\vspace{-0.5em}
We introduced a novel approach to implement dynamic real-time Ambisonics order adaptation for INMP applications. Our evaluation, conducted through subjective listening tests (MUSHRA), demonstrated that instantaneous Ambisonics order adjustments provided audio quality and spatial localization closely comparable to high-order.
Future work will explore enhancements to the adaptation mechanism by considering additional network performance indicators, e.g., latency and overhead, and different implementation strategies of the adaptation will be considered and evaluated.
%
%
\vspace{-0.5em}
\bibliographystyle{IEEEtran}
\bibliography{biblioEUSIPCO}
\end{document}